\documentclass[graybox, envcountchap]{svmult}

\usepackage{mathptmx}        % selects Times Roman as basic font
\usepackage{amsmath}
\usepackage{amssymb}
\usepackage{color}
\usepackage{helvet}          % selects Helvetica as sans-serif font
\usepackage{courier}         % selects Courier as typewriter font
\usepackage{dirtree}
\usepackage{makeidx}        % allows index generation
\usepackage{graphicx}        % standard LaTeX graphics tool
\usepackage{subfig}

\usepackage{multicol}        % used for the two-column index
\usepackage[bottom]{footmisc}% places footnotes at page bottom

\usepackage{hyperref}        %for hyperlinks
\hypersetup{colorlinks=true,urlcolor=blue}

\usepackage[misc]{ifsym}

\makeindex             % used for the subject index
\newcommand{\Msun}{M_{\odot}}

\newcommand{\kms}{\mbox{${\rm km~s^{-1}}$}}
\newcommand{\kmsMpc}{\mbox{km s$^{-1}$ Mpc$^{-1}$}}

\begin{document}

%%%%%%%%%%%%%%%%%%%%%%%%%%%%%%%%%%%%%%%%%%%%%%%%%%%%%%%%%%%%%%%%%

\title{The Hubble Constant: A Historical Review}
\titlerunning{H0: History}
\author{R. Brent Tully}
\institute{R. Brent Tully (\Letter) \at University of Hawaii, 2680 Woodlawn Dr., Honolulu, HI 96822, USA, \email{tully@ifa.hawaii.edu}}
\maketitle

\abstract{For 100 years since galaxies were found to be flying apart from each other, astronomers have been trying to determine how fast.
  The expansion, characterized by the Hubble constant, $H_0$, is confused locally by peculiar velocities caused by gravitational interactions, so observers must obtain accurate distances at significant redshifts.
Very nearby in our Galaxy, accurate distances can be determined through stellar parallaxes.
There is no good method for obtaining galaxy distances that is applicable from the near domain of stellar parallaxes to the far domain free from velocity anomalies. 
The recourse is the distance ladder involving multiple methods with overlapping domains.
Good progress is being made on this project, with satisfactory procedures and linkages identified and tested across the necessary distance range.
Best values of $H_0$ from the distance ladder lie in the range $73-75$~\kmsMpc.
On the other hand, from detailed information available from the power spectrum of fluctuations in the cosmic microwave background, coupled with constraints favoring the existence of dark energy from distant supernova measurements, there is the precise prediction that $H_0 = 67.4$ $\pm1\%$.
If it is conclusively determined that the Hubble constant is well above 70~\kmsMpc\ as indicated by distance ladder results then the current preferred $\Lambda$CDM cosmological model based on the Standard Model of particle physics may be incomplete.
There is reason for optimism that the value of the Hubble constant from distance ladder observations will be rigorously defined with $\sim 1\%$ accuracy in the near future.
}

%%%%%%%%%%%%%%%%%%%%%%%%%%%%%%%%%%%%%%%%%%%%%%%%%%%%%%%%%

\section{In the Beginning}\footnote{Chapter in book on the Hubble Constant Tension, Published 2023, Springer Nature, Singapore, Eds. E. Di Valentino \& D. Brout}
\label{sec:1}

In the first decades of the Twentieth century, evidence was accumulating that spiral nebulae lay beyond the Milky Way as ``island universes'', with discussion culminating in the Great Debate\cite{Shapley1921, Curtis1921}.
History books usually cite the convincing evidence for the extragalactic nature of spiral nebulae to the identification of cepheid variable stars in M31 and M33 by Edwin Hubble\cite{Hubble1925} in 1925.
  The relationship between pulsation period and luminosity had been established by Henrietta Leavitt\cite{Leavitt+1912} in 1912.
  Hubble determined the distances to each of M31 and M33 to be 285~kpc.
  However, three years earlier, a better distance to M31 of 450~kpc had been calculated by Ernst \"Opik\cite{Opik1922} assuming the virial theorem relation between velocity dispersion, dimension, and mass and assuming the solar relation between luminosity and mass.
  The modern value for the distance to M31 is 740~kpc and to M33 is 850~kpc.

  In the meantime, spectra of the spiral nebulae were being accumulated, mostly by Vesto Slipher\cite{Slipher1915}, and it was being remarked that almost all spectral features were displaced to longer wavelengths from their laboratory positions.
  In a seminal paper\cite{Hubble1929}, Hubble demonstrated the correlation between velocity displacement, subsequently called redshift, and galaxy distance, with the relationship between these parameters that now bears his name.
  In fact, the expansion of the Universe had been anticipated and hinted at by Georges Lema\^itre\cite{Lemaitre1927} on dynamical grounds.
  The static Universe posited by Albert Einstein\cite{Einstein1917} with his constant $\Lambda$ to offset gravity was unsatisfactorily unstable.
  The initial estimate of Hubble's constant\cite{Hubble1929} was $H_0=500$~\kmsMpc, later increased slightly by Hubble to 530~\kmsMpc\cite{Hubble1936}.
  The uncertainty was large, as pointed out by Jan Oort who favored a lower value\cite{Oort1931} and Arthur Eddington who favored a higher value\cite{Eddington1935}.
  Eddington provided a theoretical perspective, a prelude to an interplay that continues to be a bane of studies of the scale of the Universe.
  Eddington claimed that the combination of relativity theory and quantum theory gave a value of $H_0$ with the wonderful quote: {\it ``I have now been able to reach a solution which I regard as definitive.
  The result is 865 km. per sec. per megaparsec.''}

\section{Evident Problems}

If the Universe has been expanding at a uniform rate since its inception then it began at a time ago that is the inverse of the Hubble constant.
With Hubbles's initial value of 500~\kmsMpc\ the implied age is 2 billion years.
The discrete moment of the beginning of expansion has come to be called the ``Big Bang.'' 
Already by 1931, radioactive dating of geological rocks was giving ages of the Earth of $2-3$~Gyr\cite{Dalrymple1994}.
Surely, the Universe is older than the Earth.
A plausible resolution was offered by the steady state theory of Hermann Bondi, Thomas Gold, and Fred Hoyle\cite{Bondi+1948, Hoyle1948}, whereby matter is continuously created as space expands so as to keep constant conditions on average in perpetuity. 

However, evidence began to emerge that the early estimates of $H_0$ were too high.
Walter Baade made a breakthrough with the discovery of distinct luminosity differences between the brightest of young (Population I) stars and old (Population II) stars when he could resolve red giant branch stars in M31, M32, and NGC\,205 from Mt. Wilson during Los Angeles blackouts during WWII\cite{Baade1944}.
The stepping stone to the calibration of the cepheid period-luminosity relation had been through Galactic globular clusters which have cepheid-like variables.
It was recognized that the brightest resolved stars in the central region of M31 and in the elliptical companions were similar to the brightest stars in globular clusters.
It became apparent that there are two kinds of cepheid variables: Population II cepheids as found in environments of old stars and Population I cepheids found amongst young stars that are brighter at a given pulsation period by $\sim 1.5$~mag\cite{Baade1954}, implying a factor 2 increase in cosmological distances.
Baade provided a very nice history of this development in 1956\cite{Baade1956}.

\section{Grasping at Straws}

Work on the extragalactic distance scale picked up in the 1950's with the inauguration of the Hale 200-inch telescope on Mt. Palomar.
There was the important contribution by Milton Humason, Nicholas Mayall, \& Allan Sandage\cite{Humason+1956}.
Imaging of the Virgo cluster galaxy NGC\,4321 revealed that features taken earlier to be individual stars were, in fact, HI regions.
The brightest individual stars are some 1.8~mag fainter.
Humason, Mayall, and Sandage concluded $H_0 \sim 180$~\kmsMpc.
Soon after, Sandage\cite{Sandage1958} reviewed the situation.
In addition to the confusion between old and young populations with cepheids, he argued that the apparent width of the cepheid period-luminosity relation at fixed period caused a significant under-estimation of  distances.
He placed $H_0$ between 50 and 100 \kmsMpc.

Over the next two decades, there was increased interest in the distance scale but there were only marginal improvements.
The problem to be addressed was clear but the tools were blunt.
Eric Holmberg tried using a correlation between the absolute magnitude and surface brightness of galaxies\cite{Holmberg1958}.
Sydney van den Bergh produced a luminosity classification for mid- to late-type spirals that could be given an absolute calibration\cite{vandenBergh1960a, vandenBergh1960b}.
Jose Sersic promoted the use of the diameter of HII regions\cite{Sersic1960}.
Ren\'e Racine compared the globular cluster luminosity function between M31 and a galaxy in the Virgo cluster\cite{Racine1968}.
In 1970, van den Bergh summarized results from nine methods: the diameter of HII regions, luminosity classes, brightest globular clusters, assumptions regarding mass-to-light ratios, magnitude of third brightest galaxy cluster member, supernovae, a relation between the surface brightness and diameter of galaxies, and the magnitude of the brightest star in a galaxy\cite{vandenBergh1970}.
None of these methods survives as useful today.
The use of supernovae in 1970 was rudimentary.
The mass-to-light method was a variation of what was used by \"Opik\cite{Opik1922}.
It was assumed that the total mass of a galaxy, $M_t$, is given by the virial theorem, with the dependency $M_t \propto \Delta V^2 R$, where $\Delta V$ is a measure of the galaxy kinematics and $R$ is an estimate of the galaxy radius.
A relationship is assumed between mass and luminosity.
The kinematic data for this so-called indicative mass procedure were provided by Morton Roberts' observations of the linewidths of galaxies in the neutral hydrogen 21\,cm line\cite{Roberts1969} or subsequent follow up.

\section{$H_0$ 50 or 100?}

The discovery in 1965 of the permeation of space with the cosmic microwave background provided compelling evidence of a hot Big Bang beginning of the Universe\cite{Penzias+1965, Dicke+1965}, and by implication a finite age.
Necessarily, there should be compatibility between the expansion age inferred from the inverse of the Hubble constant, the ages of objects within the Universe, and a model that describes the evolution of the Universe since its inception.
Two simple models to consider are a completely empty universe or a topologically flat universe filled with matter.
With $H_0=50$~\kmsMpc\ ages would be 19.3 and 13.0~Gyr, respectively, with these models.
With $H_0=100$~\kmsMpc\ the respective ages would be 9.7 and 6.5~Gyr.

Ages of stars within our galaxy could be estimated through radioactive dating, white dwarf cooling or, particularly, the main sequence turnoff magnitude within globular clusters.
Into and through the 1980's, the best estimate of the oldest ages found in the Galaxy were $\sim 16$~Gyr\cite{Janes+1983, Chieffi+1989}, with the Universe presumably 1~Gyr or so older.
Evidently, the Universe is not empty.
Occam's razor favored models that did not invoke a cosmological constant.
An open universe with the density of matter 25\% of the critical density to achieve closure ($\Omega_m=0.25$) and $H_0=50$~\kmsMpc\ would be consistent with an age of 16~Gyr.
The age of a closed universe ($\Omega_m=1$) and $H_0=50$, at 13~Gyr, would be marginally consistent with stellar ages. 
Such simple models were inconsistent with those stellar ages if the Hubble constant was much above 50~\kmsMpc. 

Sandage and Gustav Tammann, in a series of papers between 1974 and 1982, presented arguments that favored low values of $H_0$\cite{Sandage+1974III, Sandage+1974IV, Sandage+1975V, Sandage+1975VI, Sandage+1982VIII}.
The methods they used to acquire distances included sizes of HII regions, brightest stars, luminosity classifications, and bootstraps through statistical properties of the Virgo cluster to applications in the field.
Estimates of $H_0$ in these papers ranged from 50 to 57 with typical errors $\pm 7$ or less.
G\'erard de Vaucouleurs, in a series of rebuttals between 1978 and 1986, favored high values of $H_0$\cite{deVaucouleurs+1979, deVaucouleurs+1981, deVaucouleurs1983, deVaucouleurs+1985, deVaucouleurs+1986}.
He and colleagues built a ladder founded on primary novae, cepheids, RR~Lyrae, and horizontal branch stars, secondary open and globular clusters, brightest blue and red supergiant stars, brightest HII loops or rings, and the velocity dispersionspo of HII regions, then a tertiary luminosity index calibration.
They found values of $H_0$ between 90 and 110 \kmsMpc.
So was born the $H_0$ 50 or 100 controversy.

\subsection{Better Methods}

Already by the 1970's, there was a reasonable foundation for the distance scale through parallaxes and the distances they gave to cepheids and RR Lyrae star on the horizontal branch, albeit subject to improvements.
But there had been only crude ways to estimate distances beyond the Local Group and into the domain unaffected by random motions.
It was in this time period that methodologies began to emerge that are still retained today for the measurement of galaxy distances at significant redshifts.

In 1977, this author and Richard Fisher published two papers finding $H_0=75-80$~\kmsMpc\ based on the correlation between the absolute luminosities of spiral galaxies and their rotation rates\cite{Tully+1977, Fisher+1977}, subsequently to be referred to as the TF relation.
This procedure recalls the indicative mass method but deviates from the parameterization of the virial theorem.
It invokes only two observables, a distance dependent luminosity and a distance independent measure of the rotation rate (taking account of the galaxy inclination and internal obscuration).
A power law was assumed between luminosity and rotation rate, with a free index subject to empirical definition.
There is the obvious but simple explanation for the correlation that more massive spiral galaxies tend to be more luminous and spin faster.
More massive galaxies tend to contain fractional more old stars, hence are redder. 
Therefore, it was not a surprise that the free index of the TF relation increases toward longer wavelengths\cite{Tully+1982},
The methodology was more solidly grounded in 1979 by shifting from photographic blue magnitudes to photoelectric near-infrared magnitudes, as demonstrated by Marc Aaronson, John Huchra, and Jeremy Mould\cite{Aaronson+1979}.
  These authors demonstrated that the free index would approximate four, $L \sim \Delta V^4$, if spiral galaxies share dimensionless scale-length mass profiles/rotation curves, have disks with a common central surface brightness, and have the same mean mass-to-light ratio.
This latter condition is more easily met in the infrared.
The true situation is more complicated.
Indeed, it is mainstream to assume that dark matter dominates the mass, so it could be considered a puzzle that the correlation between star light and kinematics is as tight as observed.
The interest of most theorists has been on the implications for galaxy formation and evolution\cite{Okamoto+2005, Governato+2007, Trujillo-Gomez+2011, Guedes+2011, Marinacci+2014, Glowacki+2020}.
Treating the relation as an empirical phenomenon, Aaronson and colleagues determined values of $H_0$ between 90 and 95 over the years 1980-1986\cite{Aaronson+1980, Bothun+1984, Aaronson+1986}.
An early detailed mapping of the velocity field of the Local Supercluster was carried out by this collaboration\cite{Aaronson+1982}. 

A contemporaneous discovery with the TF relation, was an analogous correlation between the luminosities and central velocity dispersions of early-type galaxies, ellipticals and lenticulars, by Sandra Faber and Robert Jackson\cite{Faber+1976}.
The two parameter formulation for ellipticals had higher dispersion than the TF relation for spirals but in 1987 it was found that the addition of distance-independent surface brightness as a third parameter provided a significant improvement\cite{Djorgovski+1987, Dressler+1987a}.
The procedure, with parameters gathered into distance dependent and independent parts, resulted in what became called the fundamental plane.
Initially there was no attempt at an absolute calibration but the method was used to map {\it relative} distances with redshifts.
A group soon known as the Seven Samurai (David Burstein, Roger Davies, Alan Dressler, Sandra Faber, Donald Lynden-Bell, Roberto Terlevich, and Gary Wegner) demonstrated the existence of a large scale flow toward a mysterious ``Great Attractor''\cite{Dressler+1987b, Lynden-Bell+1988, Burstein+1990, Dressler+1990}.
Even as late as 2001 when substantial samples had been accumulated, the fundamental plane was still only being used to derived peculiar velocities and velocity flows based on relative distance measurements\cite{Colless+2001, Hudson+2001, Bernardi+2002}.
However, it has found an important place today within absolute measurements of the extragalactic distance scale. 

A next contribution to the pantheon of important methodologies for the measurement of galaxy distances came in 1988 with recognition that distance information is encoded in the surface brightness fluctuation amplitudes of dominantly old stellar populations\cite{Tonry+1988}.
The brightest stars in regions devoid of recent star formation lie on the upper red giant branch of stellar evolution.
The relative number of such stars within an observer's pixel increases as the square of distance, causing a predictable damping with distance.
Between 1989 and 2000, John Tonry and colleagues exploited this phenomenon to derive estimates of the Hubble constant that fell from an initial $H_0=88$ to a later $H_0=77$~\kmsMpc\cite{Tonry+1989, Tonry1991, Tonry+1997, Tonry+2000}.
The study culminated in the publication of 300 distances to galaxies within 4000~\kms\ and a study of the flow patterns discerned from this sample\cite{Tonry+2001}.   

The foundation of the surface brightness fluctuation technique is the constancy of luminosities of stars at the ``tip'' of the red giant branch, where degenerate helium cores reach a critical mass for the onset of burning to carbon and subsequently a star finds itself on the horizontal branch.
The phenomenon harkens back to Baade's identification of Population~II stars in M31 and companions\cite{Baade1944}.
Jeremy Mould and others were using observations of such stars in the 1980's to estimate distances to galaxies within the Local Group\cite{Mould+1986}.
As charge coupled device (CCD) detectors became common-place, the so-called TRGB method became well established\cite{DaCosta+1990, Lee+1993}.
Observations from ground-based telescopes began to push slightly beyond the Local Group\cite{Sakai+1996}.
With space facilities the TRGB method would evolve to be an important pedestal for extragalactic distances. 

Other methods of possible merit explored in those years have not found a place as important players today.
There was evidence that the planetary nebulae luminosity function (PNLF) had small scatter\cite{Jacoby+1990, Mendez+1993} and was giving values of $H_0=86$~\kmsMpc.
However, the PNLF has shown limited usefulness.
It required an empirical calibration to a statistically significant number of galaxies, cross-referenced with fundamental methods, but once calibrated had limited range of application.
A similar criticism can be made of the use of the Balmer spectral features in very luminous B and A spectral class supergiants\cite{Tully+1984, Kudritzki+2014}.
Potentially, this Flux weighted Gravity-Luminosity Relationship (FGLR) could be very accurate but its range is limited and requires considerable effort per target.
There was once hope for the utility of the globular cluster luminosity function which gave values of $H_0$ in the range 75-80\cite{Hanes1979, McMillan+1993, Whitmore+1995}.
However it was difficult to have confidence in results when galaxy globular cluster populations are so variable and the formation processes are in dispute. 
Novae were found to give $H_0=70$~\kmsMpc\ with limited usage\cite{Pritchet+1987, dellaValle+1995}.

\subsection{Model-based Methods}

All of the methods discussed so far have been linked to the distance ladder that starts with stellar parallaxes and builds through properties of stars in various parts of the Hertzprung-Russell diagram like the main sequence turnoff, red giant and horizontal branches, and variable RR Lyrae and cepheid stars.
Further steps on the ladder can proceed by association of such features in external galaxies, providing absolute scales to the sorts of methods that have been discussed.
With exceptions yet to be discussed, these methods access limited redshifts.
There are other ways of estimating galaxy distances, though, that avoid the calibration ladder and can provide information on the expansion scale with a long reach.

One of these involves gravitational lens time delays.
Sjur Refsdal explored the concept in 1964, well before there was observational evidence of the phenomenon\cite{Refsdal1964}.
A gravitational lens caused by a mass concentration could cause light from a background source to follow multiple paths to reach an observer, with the time followed on each path slightly different.
If the emission from the background source is fluctuating, say because it is a supernova or active galactic nucleus, then the arrival to the observer of that signal will be slightly different along each of the paths.
Roughly, the time delay between two paths, $\Delta t$, is described by the formula\cite{Kochanek+2003}
\begin{eqnarray}
  \Delta t \sim 0.5 \left[ {1+z_l \over c} \right] \left[ {d_l d_s \over d_{ls}} \right] (\Theta_A^2-\Theta_B^2) 
\label{eq:deltat}
\end{eqnarray}
where $d$ are angular diameter distances, $d_l$ from observer to the lens (at redshift $z_l$), $d_s$ from observer to the source, and $d_{ls}$ from lens to source, and $\Theta_A$, $\Theta_B$ are angular separations from the lens of images A and B. 
Each of the distances depends inversely on $H_0$, permitting a reformulation of the time delay to give a value of the expansion parameter within an assumed world model.
Early estimates of $H_0$ were in the range 50 to 75 and even greater than 100, with recognized large uncertainties related to the distribution of mass around the lens\cite{Borgeest+1984, Rhee1991}.

Another model dependent methodology developed out of a phenomenon first discussed by Rashid Sunyaev and Yakov Zel'dovich\cite{Sunyaev+1972}.
The possible exploitation to obtain distances was discussed by Joe Silk and Simon White\cite{Silk+1978} and applied by Mark Birkinshaw and colleagues and others\cite{Birkinshaw+1991, Birkinshaw+1994, Herbig+1995, Rephaeli1995}.
The so-called Sunyaev-Zel'dovich (SZ) effect pertains to a dimming of the cosmic microwave background radiation in the direction of rich clusters of galaxies due to inverse-Compton scattering of the radiation by electrons in the hot cluster halo.
Thermal bremsstrahlung flux can also be observed from that same hot gas at X-ray wavelengths.
The SZ temperature decrement depends on the electron temperature and density along the line-of-sight in the cluster.
The X-ray flux depends on the square root of the electron temperature, the product of the electron and proton densities (effectively the square of electron density), the volume of the hot gas cloud, and because it is an apparent flux, inversely on the square of distance. 
If all of the microwave decrement, the X-ray flux, and the gas temperature can be measured then an estimate of distance can be derived from the observed size and shape of the cluster and a model of the gas distribution.
It is to be assumed that the gas is in hydrostatic equilibrium, which can clearly not be the case if a cluster is experiencing a merger event but might be a reasonable approximation in some cases.
There is the complication that the X-ray flux, depending on electron density squared, will particularly arise in high density regions of the gas, while the SZ effect arises from lower electron density, longer path length regions.
Complex modelling is required.
The early analyses gave values of the Hubble constant in the range 45-70~\kmsMpc.

Today supernovae play an important role in measurements of the scale of the Universe but in early years their contributions were ambiguous.
Walter Baade made clear the distinction between novae that can be recurring events in stars and much brighter supernovae, and with Fritz Zwicky, attempted to estimate their absolute magnitudes\cite{Baade1938, Baade+1938}.
The spectra of supernovae vary considerably over time within an event and between events.
There are those without, or almost without, the spectral features of hydrogen (called Type I) and those exhibiting strong hydrogen features (called Type II).
Various sub-classes can be identified by spectral or light curve characteristics.
There are two fundamentally different paths to supernova cataclysms.
There are core collapse events in massive stars that have exhausted their nuclear fuels, usually identified as Type~II although also as Type~Ib or Type~Ic.
Then there are Type~Ia merger or accretion events that take white dwarfs over the Chandrasekhar limit of $1.44~\Msun$ for electron degeneracy pressure support (or at least close enough to that limit to cause ignition). 
The Type~Ia supernovae are brighter, hence seen to greater distances and predominant in surveys because of access to greater volumes.
Charles Kowal made an early attempt to define the absolute luminosities of the various supernova classes\cite{Kowal1968}.

Attempts were made to specify absolute magnitudes using models.
In principal, a link could be made between the distance independent expansion velocity of the supernova photosphere and the distant dependent angular expansion as determined by measurements of temperature and apparent magnitude\cite{Branch+1973, Kirshner+1974, Branch+1981, Branch1987, Schmidt+1994, Fisher+1995}.
The book from a meeting in 1984 provided a good summary of the status of Hubble constant estimates at that date from supernova considerations\cite{Bartel1985}.
With 13 separate model-driven contributions published up to 1995, the average value of $H_0$ was 58~\kmsMpc.
A leader in the field who can remain anonymous once said ``Either $H_0$ equals 50 or we don't understand core-collapse supernovae.''
Right, one of those.

\section{Satellites, Supernovae, and the CMB}

Changes were coming in the 1990's.
At the foundational level, the Hipparcos satellite that operated from 1989 to 1993 greatly increased the number and improved the quality of stellar parallaxes.
Distances to globular clusters were increased and consequently their ages decreased.
By the end of the decade, it was being argued that the oldest stars in the Galaxy had ages $12-13$~Gyr\cite{Chaboyer+1998, Pont+1998,Carretta+2000}.

In 1990, Hubble Space Telescope (HST) was launched.
After a false start, a solution was implemented with a service mission and the addition of Wide Field/Planetary Camera 2 (WFPC2) in 1993.
The full HST bounty only began to be harvested, though, after the third service mission in 2002 with the activation of the much more powerful Advanced Camera for Surveys (ACS).  
Only hints of the considerable role that HST was to play were revealed in the last decade of the 20th century.
Even before the correction of the optics, two teams began using HST to observe cepheid variables in nearby galaxies, switching to the use of WFPC2 as it became available.
One team involved Abhijit Saha, Alan Sandage and collaborators\cite{Saha+1994,Saha+1996,Saha+1997,Saha+1999,Saha+2001,Sandage+1996a}.
Their results were summarized by Tamman, Sandage, and Saha in 2000 who concluded that the distances to SNIa calibrated by cepheids gave $H_0=58\pm6$~\kmsMpc\cite{Tammann+2003}.
The other team lead by Wendy Freedman was formed to carry out what became known as the HST Key Project\cite{Freedman+1994a, Freedman+1994b, Kelson+1996, Turner+1998, Gibson+2000}.
The Key Project study, cumulatively grounded in cepheid, tip of the red giant branch, globular cluster, planetary nebulae, and surface brightness fluctuation observations\cite{Ferrarese+2000}, arrived at a value of the Hubble constant of $72\pm8$~\kmsMpc\cite{Freedman+2001}.\footnote{Up to this point in the discussion, values of $H_0$ have been given without errors.  For the most part, errors given in the last century (and often the current one) have been wildly optimistic.}

The effort to calibrate the SNIa scale was two-pronged.
Paralleling the work acquiring cepheid distances to hosts of SNIa were vast improvements in SNIa acquisitions.
The utility of the SNIa tool was much improved by the adjustments that could be made to minimize scatter through the recognition by Mark Phillips that the peak brightness depends on the decay rate of luminosity after peak brightness\cite{Phillips1993}.
Diverse teams other than the those of the Key Project or Saha-Sandage that looked at what could be done with SNIa in the 1990's found values of $H_0$ between 63 and 70\cite{Riess+1995, Jha+1999, Phillips+1999, Wang2000, Hernandez+2000}.

Throughout this period, other methodologies continued to be used, generally contributing more noise than signal.
Sandage and co-workers published on studies using a spiral luminosity function, spiral lookalikes, the globular cluster luminosity function, and their own analysis of the TF relation, finding in each case $H_0=57$ within $\pm5$~\kmsMpc\cite{Sandage1996a, Sandage1996b, Sandage+1996b, Federspiel+1998}.

The most exciting developments with supernovae at this time did not directly involve the Hubble constant.
Observations of high redshift SNIa by teams lead by Brian Schmidt and Saul Perlmutter\cite{Schmidt+1998, Perlmutter+1999} provided evidence contrary to the cold dark matter (CDM) model with sufficient density in matter for a closed universe ($\Omega_m=1$).  
The most distant supernovae were too faint.
Assuming that they had intrinsic luminosities similar to nearby SNIa, the observations could best be fit with a model with cosmic acceleration, such as anticipated with the inclusion of the cosmological constant, $\Lambda$, in Einstein's equation.

In quite another domain, a satellite to study the cosmic microwave background was to have a profound impact on our understanding of the scale of the Universe.
The Cosmic Background Explorer (COBE) that operated between 1989 and 1993 demonstrated unambiguously the black body spectrum of the cosmic microwave background, strong proof for its origin in a hot Big Bang beginning of the Universe\cite{Smoot+1992, Bennett+1996}.
The strong hints of a peak in the power spectrum of fluctuations was confirmed by the balloon experiments MAXIMA and BOOMERANG\cite{Hanany+2000, deBernardis+2000, Netterfield+2002}.
The angular scale of the first peak provided compelling evidence that the Universe has a flat topology.
The results from distant SNIa indicated that cold dark matter alone could not produce a spatially flat universe and gave a preference for the existence of vacuum energy.
The microwave background and SNIa observations taken together were consistent with a $\Lambda$CDM model of a universe with both dark matter with density $\Omega_m$ and dark energy with density $\Omega_{\Lambda}$ such that $\Omega_m + \Omega_{\Lambda} = 1$. 

There was yet another ingredient in the mix.
Wide angle redshift surveys were accumulating and more power was being seen on large scales than anticipated by the standard CDM model with $\Omega_m=1$\cite{Saunders+1990, Saunders+1991, Efstathiou+1990, Loveday+1992}.
The disagreements became more accute with the additional constraints imposed by the microwave background power spectrum results from COBE\cite{Bunn+1997, Peacock1997, Bond+1999}.
Models with $\Omega_m\sim0.3$ were favored, whether with or without dark energy.
Perversely, models with $\Omega_m=1$ were still being offered that could be reconciled with the large scale structure and microwave background information but only if $H_0$ was taken to be $\sim 35$~\kmsMpc, in utter contradiction with distance ladder observations\cite{Bartlett+1995, White+1995, Lineweaver+1997}.

Increasingly, wide field imaging and redshift surveys were becoming available as the 20th century came to a close.
Michael Hudson was reconstructing density and velocity fields from these inputs\cite{Hudson1993, Hudson1994}.
It was demonstrated that the TF relation could give distances with comparable accuracy observing at optical $R$, $I$  bands as obtained in the infrared, given area photometry using CCDs\cite{Pierce+1988}.
Jeff Willick and colleagues made major contributions to issues of calibration and biasing\cite{Willick1994, Willick+1995, Strauss+1995, Willick+1996, Willick+1997a, Willick+1997b}.
They came to favor a value of the Hubble constant of 85~\kmsMpc\ and, from the peculiar velocity field,  matter density $\Omega_m \sim 0.3$\cite{Willick+1998, Willick+2001}.
Riccardo Giovanelli, Martha Haynes, and colleagues greatly enlarged the TF relation sample sizes with extensive 21cm observations with Arecibo Telescope.
They derived $H_0=69$~\kmsMpc\cite{Giovanelli+1997}.

A bit of a wild card was introduced with the introduction by Avishai Dekel and others of an analysis tool that came to be called POTENT\cite{Dekel+1990, Dekel+1993}.
A smoothed three-dimensional velocity field was inferred from observed radial velocities assuming a potential flow, consistent with peculiar velocities arising from density perturbations.
Initial conditions were linked with final positions using the Zel'dovich approximation assuming linear theory of the growth of structure.
Dekel et al. were able the generate maps of the velocity and density field within a radius of 8,000~\kms\ that impressively resemble maps being built two decades later.
However, it was a considerable distraction that values of matter density were being inferred that were $\Omega_m \sim 1$ or only slightly less\cite{Dekel+1999}.
A similar high value was found for $\Omega_m$ from a power spectrum analysis of the density fluctuations arising from the Giovanelli et al. observed peculiar velocities\cite{Freudling+1999}.
By contrast, an analysis based on numerical action orbit reconstructions by Ed Shaya, Jim Peebles, and this author derived a value of matter density of $\Omega_m = 0.17\pm0.10$\cite{Shaya+1995}.
Igor Karachentsev and others were remarking on the extremely cold local flow\cite{Karachentsev+2002}.
It could be seen from orbit reconstructions that if $\Omega_m \sim 1$ then there would be the triple-value signatures of collapse\cite{Tonry+1981} all over the map, a situation far from actuality.
The situation was nicely summarized by contributions at a workshop organized by St\'ephane Courteau in 1999 on cosmic flows in Victoria, Canada\cite{Courteau+2000}.

As the century came to an end, there was a remarkable paradigm shift.
A foundation for it had been laid 20 years earlier by Alan Guth's ideas of the origin of the Universe in an inflation driven by vacuum energy\cite{Guth1981}.
It was thought natural that this origin would result in a universe that is close to topologically flat.
The power spectrum of cosmic microwave background fluctuations with a peak on an angular scale of about a degree provided compelling evidence that this is the case.
The measurements that SNIa at $z\sim1$ are fainter than expected if $\Omega_m =1$ favored the requirement of a positive $\Lambda$ contribution.
There was general agreement that redshift surveys were indicating that there was too much structure on large scale and that biasing between the distribution of galaxies and the distribution of matter was insufficient to rescue $\Omega_m=1$ models.
Even though it was demonstrated that there are large scale flows\cite{Dressler+1987b}, flows on local scales are quiet, direct evidence for modest fluctuations in matter.
Then there was abundant evidence from measurements of the Hubble parameter that, assuredly, the CDM universe with closure density in matter was untenable and even an open universe with $\Omega_{\Lambda}=0$ was contradicted.

The one model that accommodated the observations posited that the Universe is spatially flat due to the combination of a currently dominant ``dark energy'' and sub-dominant matter, most of it ``dark matter''.
This $\Lambda$CDM model relieved the Hubble constant tension at the time.
Any value of $H_0$ in the range being proposed could be accommodated.
The value from the TF relation favored by this author as of 2000 was $H_0=77$~\kmsMpc, lower than earlier estimates\cite{Pierce+1988} primarily because of the availability of 24 zero-point calibrators, whereas there were only 3 in 1988\cite{Tully+2000}.
The HST Key Project derived the value $H_0=71$ from their calibration of the TF relation\cite{Sakai+2000}.
Overall, the Key Project best value was $H_0=72$~\kmsMpc\cite{Freedman+2001}.
For a brief period, there was no Hubble constant tension.
That situation would not last.

\section{$H_0$: 67 or 75?}

Increasingly detailed observations were made of the cosmic microwave background with the Wilkinson Microwave Anisotropy Probe (WMAP) that operated between 2001 and 2010 and with the Planck satellite between 2009 and 2013.
The spatial power spectrum of fluctuations could be fit in impressive detail with a $\Lambda$CDM model with $H_0=67.4\pm 0.5$~\kmsMpc\ and $\Omega_m = 0.315\pm 0.007$\cite{Planck2014, Planck2020, Efstathiou+2021}.
Acoustic oscillations stimulated around dark matter fluctuations in the plasma of the hot Universe would have an imprint in baryon structures in the cold Universe on the scale of the sound horizon at recombination.
These baryon acoustic oscillations (BAO) give rise to a feature in two-point correlation functions over a range of redshifts, obtained in studies of large galaxy redshift surveys like the SDSS-III Baryon Oscillation Spectroscopic Survey (BOSS) and the Dark Energy Survey (DES)\cite{Aubourg+2015, Macaulay+2019, Lemos+2019}.
The angular scale of the BAO feature closely tracks the evolution of the Hubble parameter as a function of redshift, $H(z)$, as anticipated by the $\Lambda$CDM model defined by the Planck analysis.
Similar consistent results are found from the power spectrum in the galaxy distributions of these redshift surveys\cite{dAmico+2020}.
These exceptionally stringent observational constraints require that $H_0 \simeq 67$~\kmsMpc\ within $\sim 1\%$ if the popular $\Lambda$CDM model correctly describes our Universe.

Meanwhile since 2000, attention has been given to the various facets of the distance scale ladder and significantly higher values of $H_0$ were reported, giving rise to a new Hubble tension.
Methodologies have been diverse and details regarding each will be discussed in the ensuing chapters in this book.
Only highlights of major programs will be given attention here.

Begin at the foundation with HST.
Though the number of targets is yet small, high quality stellar parallaxes have been obtained with HST to Galactic cepheids through a spatial scanning technique\cite{Riess+2018}.
The observations are important because, earlier, all but two of the Galactic cepheids that had good parallax distances have shorter periods ($T<10$ days) than the cepheids studied in external galaxies. 
There are to date HST observations of cepheid variables in 42 host galaxies of SNIa events.  The coupling with SNIa observations out to $z\sim 1$ yields $H_0 = 73.3\pm1.0$~\kmsMpc\cite{Riess+2022a}.

Taken at face value, the difference between this ladder $H_0$ and the early universe $\Lambda$CDM model $H_0$ is $5\sigma$. 
The concern is unidentified systematics.
For confirmation, a completely independent path is needed.
Such a path is being explored, although it is not yet fully developed.
It begins with the replacement of cepheids for distance measurements to nearby galaxies with tip of the red giant branch.
Currently, TRGB absolute magnitudes are established through parallax measurements to horizontal branch and RR Lyrae stars, giving distances to Galactic clusters and Milky Way companion satellites, whence establishing the absolute magnitudes of the brightest stars on the red giant branch.\cite{Carretta+2000, Rizzi+2007}
The astronomy community was startled in 2020 when Wendy Freedman and team replaced the calibration of SNIa with TRGB distances to host galaxies in place of cepheid distances, and found $H_0=69.6\pm1.7$~\kmsMpc, consistent with the early universe value\cite{Freedman+2020}.
A re-evaluation of the study determined $H_0=71.5\pm1.8$~\kmsMpc, consistent but leaning toward tension\cite{Anand+2022}.

To have a completely separate path, there needs to be an alternative to SNIa that can be applied at substantial redshifts.
Core collapse supernovae (SNII) offer a possibility.
It has been demonstrated that there is a correlation between expansion velocities and peak magnitudes, with scatter of about 15\%.
With still a small sample, a calibration based on cepheids is giving $H_0=76\pm5$~\kmsMpc\cite{deJaeger+2020}.
The surface brightness fluctuation technique probably offers an even greater avenue for progress because observations in the infrared with space telescopes give distances with accuracies of 5\% and can reach to several hundred megaparsecs.
With current applications, again grounded through cepheid calibrations, $H_0 = 73.3\pm2.4$~\kmsMpc\ is being found\cite{Blakeslee+2021}.

Programs to study the velocity field of galaxies have been concurrent with efforts to define the Hubble constant.
The peculiar velocity of a galaxy can be obtained independent of an absolute calibration and hence of $H_0$.
Very large samples have been accumulated involving the TF and fundamental plane relations, extending to $0.05c-0.1c$ and providing wide coverage of the sky.
Individual distance uncertainties are $20-25\%$ but sample sizes are in the 10's of thousands.
There have been two noteworthy fundamental plane contributions arising from wide field imaging and redshift programs: the Six Degree Field Galaxy Survey for peculiar velocities (6dFGSv) targeting 9,000 E$-$S0 galaxies in the south celestial hemisphere out to 16,000~\kms\cite{Springob+2014, Qin+2018}, and the SDSS peculiar velocity sample of 34,000 E$-$S0 galaxies in the north celestial, north Galactic quadrant extending to 30,000~\kms\cite{Howlett+2022}.
The latter sample was given a zero-point calibration ultimately based on cepheids, which resulted in the determination $H_0 = 75\pm2$~\kmsMpc.

Complementary observations of spiral galaxies have received impetus from extensive 21cm neutral hydrogen survey programs\cite{Courtois+2015, Haynes+2018}.
Major velocity field studies have been published based on TF relation distance measurements\cite{Springob+2007, Tully+2008, Tully+2013, Tully+2016, Hong+2019}.
It has been suggested that the correlation with spiral galaxy rotation rate is tightened if HI flux is added to the optical or infrared flux, creating a measure of the total baryonic mass.
The resultant correlation has been called the baryonic TF relation (BTFR)\cite{McGaugh+2000, Lelli+2016}.
The BTFR has been used with a sample of 10,000 galaxies, with zero point scaling set by a combination of cepheid and TRGB distances, to obtain $H_0 = 75.5\pm2.5$~\kmsMpc\cite{Kourkchi+2022}.

It is an unfortunate feature of the distance ladder that there is no one methodology that is useful from the nearby within the range of parallax measurements to the redshift regime beyond peculiar velocity effects.
By necessity, methodologies must be interlinked.
If the Hubble constant is to be determined robustly from the ladder, the linkages must be robust, which is to say, they must be abundant and diverse.
However, making linkages is tricky.
For example, cepheid variables are only found in systems with young stars; likewise SNII.
The use of the TF relation is restricted to spiral galaxies.
The fundamental plane and surface brightness fluctuation methods are only used with massive early-type galaxies.
The TRGB method can be used to derive distances to essentially any galaxy but only if it is nearby; within 10 Mpc observing with HST, unless a special effort is made.
SNIa can arise in any kind of host but they are rare nearby.
There is the probability that there are subtle variations in the luminosities of SNIa that depend on host properties, such as whether or not there are young stars, which raises issues if calibrations only involve cepheids\cite{Hayden+2013, Childress+2013, Roman+2018}.

By necessity, unless targets are within the reach of cepheid or TRGB calibration, linkages between methods must be through group affiliations.
Groups can range from clusters of many hundreds of members to pairs.
Linkages between methodologies across this full range are useful and desirable.
Good group catalogs should have a physical basis, have a high level of completion among the galaxy samples with distances, and have minimal contamination from interlopers.
Groups considered appropriate\cite{Tully2015, Kourkchi+2017, Tempel+2017} were used to acquire linkages between the largest available samples of the procedures described above in the compilation of the Cosmicflows-4 compendium of 56,000 galaxy distances\cite{Tully+2023}. 
Samples were brought to a common absolute scale based on cepheid and TRGB calibrations with a Bayesian Markov chain Monte Carlo (MCMC) analysis (where a small cepheid-TRGB scale difference was reconciled with preliminary parallax information from Gaia satellite observations).
The value of the Hubble constant from the Cosmicflows-4 compilation is arguably the best available determination from the distance ladder because it integrates all the major methodologies and surveys.
$H_0 = 74.6\pm0.8$~\kmsMpc\ is reported\cite{Tully+2023}.
This statistical error is very small because the combined sample of 56,000 galaxies in 38,000 groups is so large.
It is estimated that systematic errors could be as large as 3~\kmsMpc.

\subsection{$H_0$ from Models}

A number of modeled observations that are independent of the distance ladder have been proposed that have dependence on the Hubble constant, several of which are discussed in this volume.
In each instance, it must be asked if the model constrains the Hubble constant or if the Hubble constant constrains the model.
To this reviewer's awareness, the strongest case of a model constraint on a galaxy distance is provided by the nuclear maser observations of NGC~4258\cite{Reid+2019}, because of the remarkably fortuitous edge-on geometry, locations, and strength of the maser signals.
The occurrence of this event so near by (7.58~Mpc) is one of extreme luck because the few other nuclear maser cases that have been studied give distance results that are far less certain\cite{Pesce+2020}.

The detached eclipsing binary measurements that give a distance to the Large Magellanic Cloud might likewise pass the bar of informing the extragalactic distance scale\cite{Pietrzynski+2019}.
It does depend on knowledge of a surface brightness-color relation and, more, on the locations of targets across a 5~degree field (a projection that is 9\% of the LMC distance).
The claimed precision of the LMC distance is 1\% but what does that mean for an irregular system that is so large on the sky and presumably in depth?
The LMC has played an important role in the calibration of the cepheid scale but, in the fullness of time, perhaps the two Magellanic clouds will be by-passed in distance scale studies.

There are a number of physical properties of the Universe that, to the degree that they are properly modeled, bear on the value of the Hubble constant.
There are strong gravitational lensing time delays of AGN signal variations through cluster halos\cite{Birrer+2022}, gravitational waves arising from neutron star mergers\cite{Abbott+2021}, the Sunyaev-Zel'dovich effect involving comparisons of the X-ray flux and microwave background damping associated with cluster plasmas\cite{Kozmanyan+2019}, cosmic chronometers giving record to the relative passage of time with massive, passively evolving galaxies\cite{Moresco+2018}, and the pair-wise galaxy separation mapping of baryon acoustic oscillation features at various epochs\cite{Alam+2021}. 
Generally, the evaluation of these interesting phenomena is done in the context of the $\Lambda$CDM paradigm, which is at the root of the $H_0$ tension controversy.

Some time ago, George Efstathiou initiated an attempt to reconcile the distance ladder measurements with the early universe implications for $H_0$\cite{Efstathiou2020}.
He explored possibilities of systematics that might affect ladder measurements; in other words, relieving the Hubble tension by postulating that the ladder distances are in error.
Of course, systematic errors are possible.
However, for this reviewer, the effort of reconciling distance scale observations with a favored world model too easily brings back memories of the end of the last century.
The young turks of the time, of which I suppose I was one, were rather convinced that, within a Friedmann-Lema\^itre-Robertson-Walker framework, only a $\Lambda \neq 0$ cosmology model would accommodate the distance scale measurements that used quality tools. 
The theoretical community too easily discounted information on the distance scale, failing to recognize that the influential voices of Alan Sandage and Gustav Tammann were grounded in the same exercise of that of Efstathiou.

\section{Future History}

Readers can be forgiven for despairing of a resolution of the question of the current expansion rate of the Universe after a hundred years of effort, though the problem is well posed. 
Happily, the tools are at hand for an unambiguous resolution with the Global Astrometric Interferometer for Astrophysics (Gaia) and the James Webb Space Telescope (JWST).
The problem can be separated into three parts: (1) the firm calibration of the methods founded within the Milky Way, specifically cepheids and TRGB, (2) the transfer of the calibration to methods that can be used at substantial redshifts, particularly SNIa and surface brightness fluctuation, and (3) the acquisition of large samples with those latter methods.

It can be anticipated that there will be serious advances on the first problem over the next year or so.
The Gaia third data release (DR3) in June, 2022, provides parallax information on thousands of cepheids, tens of thousands of RR Lyrae stars, and hundreds of thousands of red giant branch stars with corollary metallicity information.
Anything written today may be well out of date to the reader.

The way forward with problems (2) and (3) will involve paths that are independent yet connected.
There are two obvious, well-defined paths; one involving young Population I stars and the other involving old Population II stars.
The Pop~I path is already well known: cepheid variables calibrate SNIa which in turn are observed to high redshift.
The Pop~II path uses TRGB to calibrate surface brightness fluctuations in massive elliptical galaxies which in turn are observed to high redshift.
These two paths can stand alone, but there are ample opportunities for cross-links through commonality of calibrator hosts, SNIa in ellipticals, and group associations.
TRGB can directly calibrate SNIa in late as well as early galaxy types.

JWST is particularly well suited to exploit the Pop~II path.
The volume of space opened up for TRGB observations is an order of magnitude greater with JWST than with HST (from 10 Mpc to greater than 20 Mpc).
The increase in volume for surface brightness fluctuation measurements is at least 30-fold (from 100 Mpc to $\sim 300$~Mpc).
If $H_0$ is to be determined to a precision of 1\%, calibration samples must have sizes of at least $10^2$ and samples at high redshift should be significantly larger.
With JWST, it is more feasible to expect that the necessary large calibrator samples can be acquired for TRGB distances rather than for cepheid distances.

There is the prospect that the SNIa tool can be improved.
While there is excitement that the Vera C. Rubin Observatory will make acquisitions of large numbers of SNIa at large redshifts, advancing our knowledge regarding cosmic acceleration, smaller telescopes already in operation are sufficient for the simpler problem of the Hubble constant.
Surveys with names like Pan-STARRS, ATLAS, ZTF, and ASAS-SN are picking up several SNIa events within $0.1c$ every night.
If only there are the resources for manpower, samples of $10^4+$ could be gathered in a few years.
With such large samples, complemented by multi-epoch spectroscopy and photometry at multiple optical and infrared bands, the information will be available to fine-tune the SNIa distance tool technology to account for variations in the paths to supernova events.

Two other large surveys in Australia will have major impacts on studies of the nature of large scale structure and how it formed.
WALLABY is a 21cm blind survey with the Australian Square Kilometer Array Pathfinder that is expected to provide some 200,000 TF measures with a mean redshift of $z=0.05$.
The Taipan Galaxy Survey uses a multiplexed spectrograph on the UK Schmidt Telescope at Siding Springs Observatory and is anticipated to obtain some 50,000 fundamental plane distances within $z=0.1$.
These studies promise to make valuable contributions toward a determination of the Hubble constant.
However, their uncertainties of 25\% per measurement will leave them more vulnerable to systematics than can be expected with SNIa or surface brightness fluctuation measurements with 5\% individual uncertainties.

As the 20th century came to an end, ladder measurements of the Hubble constant were at odds with the favored cosmological model of the time of cold dark matter with $\Lambda=0$.
The new favorite became the $\Lambda$CDM model with dark energy giving rise to acceleration of space in a topologically flat universe.
Yet ladder measurements, continuously improving, create doubts that this currently favorite model is complete.
Yes, there is a Hubble tension.

\begin{acknowledgement}    
Two of the leaders in the quest to measure the Hubble constant, Marc Aaronson and Jeff Willick, died in tragic accidents in their prime.There are so many others who have contributed, too many to be individually acknowledged so I won't even try.
\end{acknowledgement}

%\biblstarthook{References should be \textit{cited} in the text by number.\footnote{Please make sure that all references from the list are cited in the text. Those not cited should be moved to a separate \textit{Further Reading} section.} The reference list should be \textit{sorted} in alphabetical order. If there are several works by the same author, the following order should be used: 
%\begin{enumerate}
%\item all works by the author alone, ordered chronologically by year of publication
%\item all works by the author with a coauthor, ordered alphabetically by coauthor
%\item all works by the author with several coauthors, ordered chronologically by year of publication.
%\end{enumerate}
%For the reference style, we suggest to use \textit{LaTeX (US)} from INSPIRE.}

%%%%%%%%%%%%%%%%%%%%%%%%%%%%%%%%%%%%%%%%%%%%%%%%%%%%%%%%%


\begin{thebibliography}{99}

\bibitem{Opik1922}
E.~Opik,
%        title = ``{An estimate of the distance of the Andromeda Nebula.}'',
ApJ,\textbf{55}, 406 (1922)
doi = {10.1086/142680}

\bibitem{Shapley1921}
H.~Shapley, 
Bull. Nat. Research Council \textbf{2}, 171, (1921)

\bibitem{Curtis1921}
H.~.D.~Curtis,
Bull. Nat. Research Council \textbf{2}, 194, (1921)

\bibitem{Hubble1925}
E.~P.~Hubble,
The Observatory \textbf{48}, 139, (1925)

\bibitem{Hubble1929}
  E.~P.~Hubble,
  Proc. Nat. Acad. Sci. \textbf{15}, 168, (1929)
  doi = {10.1073/pnas.15.3.168}

\bibitem{Lemaitre1927}
  G. Lema\^itre,
  Ann. Soc. Sci. Bruxelles \textbf{47}, 49, (1927)

\bibitem{Slipher1915}
  V.~M.~Slipher,
  Pop. Astron. \textbf{23}, 21, (1915)

%\bibitem{Stromberg1925}
%  G.~Stromberg,
%  ApJ \textbf{61}, 353, (1925)
%  doi = {10.1086/142898}

\bibitem{Leavitt+1912}
  H.~S.~Leavitt \& E.~C.~Pickering,
  Harvard Coll.Obs. Circ. \textbf{173}, 1, (1912)

\bibitem{Einstein1917}
  A.~Einstein,
  Sitzungsb. K\"onig. Preuss. Akad. 142, (1917)

\bibitem{Oort1931}
  J.~H.~Oort,
  BAN \textbf{6}, 155, (1931)

\bibitem{Eddington1935}
  A.~S.~Eddington,
  MNRAS \textbf{95}, 636, (1935)
  doi = {10.1093/mnras/95.8.636}

\bibitem{Hubble1936}
  E.~P.~Hubble,
  Realm of the Nebulae, Yale U.~P.
  ISBN 9780300025002
  
\bibitem{Bondi+1948}
  H.~Bondi \& T.~Gold,
  MNRAS \textbf{108}, 252, (1948)
  doi = {10.1093/mnras/108.3.252}

\bibitem{Hoyle1948}
  F.~Hoyle,
  MNRAS \textbf{108}, 372, (1948) 
  doi = {10.1093/mnras/108.5.372}

\bibitem{Baade1944}
  W.~Baade,
  ApJ \textbf{100}, 137, (1944)
  doi = {10.1086/144650}

\bibitem{Baade1956}
  W.~Baade,
  PASP \textbf{68}, 5, (1956)
  doi = {10.1086/126870}

\bibitem{Baade1954}
  W.~Baade,
  Trans. I.A.U. \textbf{8}, 397 (1954)

\bibitem{Dalrymple1994}
  G.~B.~Dalrymple,
  {\it The Age of the Earth}, Stanford U.~P.
  ISBN 0804723311

\bibitem{Humason+1956}
  M.~L.~Humason, N.~U.~Mayall, \& A.~R.~Sandage,
  AJ \textbf{61}, 97, (1956)
  doi = {10.1086/107297}

\bibitem{Sandage1958}
  A.~R.~Sandage,
  ApJ \textbf{127}, 513, (1958)
  doi = {10.1086/146483}

\bibitem{Holmberg1958}
  E.~Holmberg,
  Lund Medd. Astron. Obs. Ser. II, 136, (1958)

\bibitem{vandenBergh1960a}
  S.~van den Bergh,
  Zf. Ap \textbf{49}, 198, (1960)

\bibitem{vandenBergh1960b}
  S.~van den Bergh,
  JRASC \textbf{54}, 49, (1960)

\bibitem{Sersic1960}
  J.~L.~Sersic,
  Zf Ap. \textbf{50}, 168, (1960)

\bibitem{Racine1968}
  R.~Racine,
  JRASC \textbf{62}, 367, (1968)

\bibitem{vandenBergh1970}
  S.~van den Bergh,
  Nature \textbf{225}, 503, (1970)
  doi = {10.1038/225503a0}

%\bibitem{deVaucouleurs+1979}
%  G.~de Vaucouleurs \& G.~Bollinger,
%  ApJ \textbf{233}, 433, (1979)
%  doi = {10.1086/157405}

%\bibitem{deVaucouleurs1964}
%  G.~de Vaucouleurs,
%  AJ \textbf{69}, 737, (1964)
%  doi = {10.1086/109349}
  
%\bibitem{deVaucouleurs1993}
%  G.~de Vaucouleurs,
%  ApJ \textbf{415}, 10, (1993)
%  doi = {10.1086/173138}

%\bibitem{Sandage1962}
%  A.~R.~Sandage,
%  IAU Symp. 15, 359, (1962)

%\bibitem{Sandage1968}
%  A.~R.~Sandage,
%  ApJL \textbf{152}, L149, (1968)
%  doi = {10.1086/180201}

\bibitem{Roberts1969}
  M.~S.~Roberts,
  AJ \textbf{74}, 859, (1969)
  doi = {10.1086/110874}

\bibitem{Janes+1983}
  K.~Janes \& P.~Demarque,
  ApJ \textbf{264}, 206, (1983)
  doi = {10.1086/160587}

\bibitem{Chieffi+1989}
  A.~Chieffi \& O.~Straniero,
  ApJS \textbf{71}, 47, (1989)
  doi = {10.1086/191364}

\bibitem{Penzias+1965}
  A.~A.~Penzias \& R.W. Wilson,
  ApJ \textbf{142}, 419, (1965)
  doi = {10.1086/148307}

\bibitem{Dicke+1965}
  R.~H.~Dicke et al.,
  ApJ \textbf{142}, 414, (1965)
  doi = {10.1086/148306}

\bibitem{Sandage+1974III}
  A.~R.~Sandage \& G.~A.~Tammann,
  ApJ \textbf{194}, 223, (1974)
  doi = {10.1086/153238}

\bibitem{Sandage+1974IV}
  A.~R.~Sandage \& G.~A.~Tammann,
  ApJ \textbf{194}, 559, (1974)
  doi = {10.1086/153275}

\bibitem{Sandage+1975V}
  A.~R.~Sandage \& G.~A.~Tammann,
  ApJ \textbf{196}, 313, (1975)
  doi = {10.1086/153413}

\bibitem{Sandage+1975VI}
  A.~R.~Sandage \& G.~A.~Tammann,
  ApJ \textbf{197}, 265, (1975)
  doi = {10.1086/153510}

%\bibitem{Sandage+1976VII}
%  A.~R.~Sandage \& G.~A.~Tammann,
%  ApJ \textbf{210}, 7, (1976)
%  doi = {10.1086/154798}

\bibitem{Sandage+1982VIII}
  A.~R.~Sandage \& G.~A.~Tammann,
  ApJ \textbf{256}, 339, (1982)
  doi = {10.1086/159911}

\bibitem{deVaucouleurs+1979}
  G.~de Vaucouleurs \& G.~Bollinger,
  ApJ \textbf{233}, 433, (1979)
  doi = {10.1086/157405}

\bibitem{deVaucouleurs+1981}
  G.~de Vaucouleurs \& W.~L.~Peters,
  ApJ \textbf{248}, 395, (1981)
  doi = {10.1086/159165}
  
\bibitem{deVaucouleurs1983}
  G.~de Vaucouleurs,
  ApJ \textbf{268}, 468, (1983)
  doi = {10.1086/160972}
  
\bibitem{deVaucouleurs+1985}
  G.~de Vaucouleurs \& H.~G.~Jr.~Corwin,
  ApJ \textbf{297}, 23, (1985)
  doi = {10.1086/163499}
  
\bibitem{deVaucouleurs+1986}
  G.~de Vaucouleurs \& W.~L.~Peters,
  ApJ \textbf{303}, 19, (1986)
  doi = {10.1086/164048}
  
%\bibitem{deVaucouleurs1993}
%  G.~de Vaucouleurs,
%  ApJ \textbf{415}, 10, (1993)
%  doi = {10.1086/173138}
  
\bibitem{Tully+1977}
  R.~B.~Tully \& J.~R.~Fisher,
  A\&A \textbf{54}, 66, (1977)

\bibitem{Fisher+1977}
  J.~R.~Fisher \& R.~B.~Tully,
  ComAp \textbf{7}, 85, (1977)
  
\bibitem{Tully+1982}
  R.~B.~Tully, J.~R.~Mould, \& M.~Aaronson,
  ApJ \textbf{257}, 527, (1982)
  doi = {10.1086/160009}

\bibitem{Aaronson+1979}
  M.~Aaronson, J.~P.~Huchra, \& J.~R.~Mould,
  ApJ \textbf{229}, 1, (1979)
  doi = {10.1086/156923}

\bibitem{Aaronson+1980}
  M.~Aaronson et al.,
  ApJ \textbf{239}, 12, (1980)
  doi = {10.1086/158084}

\bibitem{Aaronson+1982}
  M.~Aaronson et al.,
  ApJ \textbf{258}, 64, (1982)
  doi = {10.1086/160053}

\bibitem{Bothun+1984}
  G.~D.~Bothun et al.,
  ApJ \textbf{278}, 475, (1984)
  doi = {10.1086/161814}

\bibitem{Aaronson+1986}
  M.~Aaronson et al.,
  ApJ \textbf{302}, 536, (1986)
  doi = {10.1086/164014}

\bibitem{Okamoto+2005}
  T.~Okamoto, V.~R.~Eke, C.~S.~Frenk, A.~Jenkins,
  MNRAS \textbf{363}, 1299, (2005)
  doi = {10.1111/j.1365-2966.2005.09525.x}
  
\bibitem{Governato+2007}
  F.~Governato et al.,
  MNRAS \textbf{374}, 1479, (2007)
  doi = {10.1111/j.1365-2966.2006.11266.x}
  
\bibitem{Trujillo-Gomez+2011}
  S.~Trujillo-Gomez, A.~Klypin, J.~Primack, A.~J.~Romanowsky,
  ApJ \textbf{742}, 16, (2011)
  doi = {10.1088/0004-637X/742/1/16}

\bibitem{Guedes+2011}
  J.~Guedes, S.~Callegari, P.~Madau, L.~Mayer,
  ApJ \textbf{742}, 76, (2011)
  doi = {10.1088/0004-637X/742/2/76}

\bibitem{Marinacci+2014}
  F.~Marinacci, R.~Pakmor, \& V.~Springel,
  MNRAS \textbf{437}, 1750, (2014)
  doi = {10.1093/mnras/stt2003}

\bibitem{Glowacki+2020}
  M.~Glowacki, E.~Elson, \& R.~Dav\'e,
  MNRAS \textbf{498}, 3687, (2020)
  doi = {10.1093/mnras/staa2616}

\bibitem{Faber+1976}
  S.~M.~Faber \& R.~E.~Jackson,
  ApJ \textbf{204}, 668, (1976)
  doi = {10.1086/154215}

\bibitem{Djorgovski+1987}
  S.~Djorgovski \& M.~Davis,
  ApJ \textbf{313}, 59 (1987)
  doi = {10.1086/164948}

\bibitem{Dressler+1987a}
  A.~Dressler et al.,
  ApJ \textbf{313}, 42, (1987)
  doi = {10.1086/164947}

\bibitem{Dressler+1987b}
  A.~Dressler et al.,
  ApJ \textbf{313}, L37, (1987)
  doi = {10.1086/184827}

\bibitem{Lynden-Bell+1988}
  D.~Lynden-Bell et al.,
  ApJ \textbf{326}, 19, (1988)
  doi = {10.1086/166066}

\bibitem{Burstein+1990}
  D.~Burstein, S.~M.~Faber, A.~Dressler,
  ApJ \textbf{354}, 18, (1990)
  doi = {10.1086/168664}

\bibitem{Dressler+1990}
  A.~Dressler \& S.~M.~Faber,
  ApJ \textbf{354}, L45, (1990)
  doi = {10.1086/164947}

\bibitem{Colless+2001}
  M.~Colless et al.
  MNRAS \textbf{321}, 277, (2001)
  doi = {10.1046/j.1365-8711.2001.04044.x}

\bibitem{Hudson+2001}
  M.~J.~Hudson et al.
  MNRAS \textbf{327}, 265, (2001)
  doi = {10.1046/j.1365-8711.2001.04786.x}

\bibitem{Bernardi+2002}
  M.~Bernardi et al.
  AJ \textbf{123}, 2990, (2002)
  doi = {10.1086/340463}

\bibitem{Tonry+1988}
  J.~L.~Tonry \& D.~P.~Schneider,
  AJ \textbf{96}, 807, (1988)
  doi = {10.1086/114847}
  
\bibitem{Tonry+1989}
  J.~L.~Tonry, E.~A.~Ajhar, G.~.A.~Luppino, 
  ApJ \textbf{346}, 57, (1989)
  doi = {10.1086/185578}
  
\bibitem{Tonry1991}
  J.~L.~Tonry,
  ApJ \textbf{373}, L1, (1991)
  doi = {10.1086/186037}
  
\bibitem{Tonry+1997}
  J.~L.~Tonry, J.~P.~Blakeslee, E.~A.~Ajhar, A.~Dressler,
  ApJ \textbf{475}, 399, (1997)
  doi = {10.1086/303576}
  
\bibitem{Tonry+2000}
  J.~L.~Tonry, J.~P.~Blakeslee, E.~A.~Ajhar, A.~Dressler,
  ApJ \textbf{530}, 625, (2000)
  doi = {10.1086/308409}
  
\bibitem{Tonry+2001}
  J.~L.~Tonry et al.,
  ApJ \textbf{546}, 681, (2001)
  doi = {10.1086/318301}

\bibitem{Mould+1986}
  J.~R.~Mould \& J.~Kristian,
  ApJ \textbf{305}, 591, (1986)
  doi = {10.1086/164273}

\bibitem{Lee+1993}
  M.~G.~Lee, W.~L.~Freedman, B.~F.~Madore,
  ApJ \textbf{417}, 553, (1993)
  doi = {10.1086/173334}

\bibitem{DaCosta+1990}
  G.~S.~Da Costa \& T.~E.~Armandroff,
  AJ \textbf{100}, 162, (1990)
  doi = {10.1086/115500}

\bibitem{Sakai+1996}
  S.~Sakai, B.~F.~Madore, W.~L.~Freedman,
  ApJ \textbf{461}, 713, (1996)
  doi = {10.1086/177096}

\bibitem{Jacoby+1990}
  G.~H.~Jacoby, R.~Ciardullo, H.~C.~Ford,
  ApJ \textbf{356}, 332, (1990)
  doi = {10.1086/168843}
  
\bibitem{Mendez+1993}
  R.~H.~Mendez, R.~P.~Kudritzki, R.~Ciardullo, G.~H.~Jacoby,
  A\&A \textbf{275}, 534, (1993)

\bibitem{Tully+1984}
  R.~B.~Tully \& S.~C.~Wolff,
  ApJ \textbf{281}, 67, (1984)
  doi = {10.1086/162075}

\bibitem{Kudritzki+2014}
  R.-P.~Kudritzki et al.
  ApJ \textbf{788}, 56, (2014)
  doi = {10.1088/0004-637X/788/1/56}
  
\bibitem{Hanes1979}
  D.~A.~Hanes,
  MNRAS \textbf{188}, 901, (1979)
  doi = {10.1093/mnras/188.4.901}
  
\bibitem{McMillan+1993}
  R.~McMillan, R.~Ciardullo, G.~H.~Jacoby,
  ApJ \textbf{416}, 62, (1993)
  doi = {10.1086/173215}
  
\bibitem{Whitmore+1995}
  B.~C.~Whitmore et al.,
  ApJ \textbf{454}, L73, (1995)
  doi = {10.1086/309788}
  
\bibitem{Pritchet+1987}
  C.~J.~Pritchet \& S.~van den Bergh,
  ApJ \textbf{318}, 507, (1987)
  doi = {10.1086/165387}
  
\bibitem{dellaValle+1995}
  M.~della Valle \& M.~Livio,
  ApJ \textbf{452}, 704, (1995)
  doi = {10.1086/176342}

\bibitem{Refsdal1964}
  S.~Refsdal,
  MNRAS \textbf{128}, 307, (1964)
  doi = {10.1093/mnras/128.4.307}

\bibitem{Borgeest+1984}
  U.~Borgeest \& S.~Refsdal,
  A\&A \textbf{141}, 318, (1984)

\bibitem{Rhee1991}
  G.~Rhee,
  Nature \textbf{350}, 211, (1991)
  doi = {10.1038/350211a0}

\bibitem{Kochanek+2003}
  C.~S.~Kochanek \& P.~L.~Schechter,
  arXiv:astro-ph/0306040, (2003)
  doi = {10.48550/arXiv.astro-ph/0306040}

\bibitem{Sunyaev+1972}
  R.~A.~Sunyaev \& Ya.~B.~Zel'dovich,
  Comm. Ap. Space Sci. \textbf{4}, 173 (1972)

\bibitem{Silk+1978}
  J.~Silk \& S.~D.~M.~White,
  ApJ \textbf{226}, L103, (1978)
  doi = {10.1086/182841}
  
\bibitem{Birkinshaw+1991}
  M.~Birkinshaw, J.~P.~Hughes, K.~A.~Arnaud,
  ApJ \textbf{379}, 466, (1991),
  doi = {10.1086/170522}
  
\bibitem{Birkinshaw+1994}
  M.~Birkinshaw, \& J.~P.~Hughes,
  ApJ \textbf{420}, 33, (1994),
  doi = {10.1086/173540}
  
\bibitem{Herbig+1995}
  T.~Herbig, C.~R.~Lawrence, A.~C.~S.~Readhead, S.~Gulkis,
  ApJ \textbf{449}, L5, (1995),
  doi = {10.1086/309616}
  
\bibitem{Rephaeli1995}
  Y.~Rephaeli,
  ARA\&A \textbf{33}, 541 (1995),
  doi = {10.1146/annurev.aa.33.090195.002545}
  
\bibitem{Baade1938}
  W.~Baade,
  ApJ \textbf{88}, 285, (1938),
  doi = {10.1086/143983}
  
\bibitem{Baade+1938}
  W.~Baade \& F.~Zwicky,
  ApJ \textbf{88}, 411, (1938),
  doi = {10.1086/143996}

\bibitem{Kowal1968}
  C.~T.~Kowal,
  AJ \textbf{73}, 1021, (1968)
  doi = {10.1086/110763}  
  
\bibitem{Branch+1973}
  D.~Branch \& B.~Patchett,
  MNRAS \textbf{161}, 71, (1973),
  doi = {10.1093/mnras/161.1.71}
  
\bibitem{Kirshner+1974}
  R.~P.~Kirshner \& J.~Kwan,
  ApJ \textbf{193}, 27, (1974),
  doi = {10.1086/153123}
  
\bibitem{Bartel1985}
  N.~Bartel,
  Lec. Not. Phys. \textbf{224}, (1985)
  doi = {10.1007/3-540-15206-7}
  
\bibitem{Branch+1981}
  D.~Branch et al.,
  ApJ \textbf{244}, 780, (1981),
  doi = {10.1086/158755}
  
\bibitem{Branch1987}
  D.~Branch,
  ApJ \textbf{320}, L23, (1987),
  doi = {10.1086/184970}
  
%\bibitem{Sandage+1992}
%  A.~R.~Sandage et al.,
%  ApJ \textbf{401}, L7, (1992),
%  doi = {10.1086/186657}
  
%\bibitem{Saha+1995}
%  A.~Saha et al.,
%  ApJ \textbf{438}, 8, (1995),
%  doi = {10.1086/175050}
  
\bibitem{Schmidt+1994}
  B.~P.~Schmidt et al.,
  ApJ \textbf{432}, 42, (1994),
  doi = {10.1086/174546}
  
\bibitem{Fisher+1995}
  A.~Fisher, D.~Branch, P.~Hoflich, A.~Khokhlov,
  ApJ \textbf{447}, L73, (1995),
  doi = {10.1086/309563}
  
\bibitem{Riess+1995}
  A.~G.~Riess, W.~P.~Press, R.~P.~Kirshner,
  ApJ \textbf{438}, L17, (1995),
  doi = {10.1086/187704}
  
\bibitem{Guth1981}
  A.~H.~Guth,
  PhRvD \textbf{23}, 347, (1981),
  doi = {10.1103/PhysRevD.23.347}
  
\bibitem{Chaboyer+1998}
  B.~Chaboyer et al.
  ApJ \textbf{494}, 96, (1998),
  doi = {10.1086/305201}
  
\bibitem{Pont+1998}
  F.~Pont, M.~Mayor, C.~Turon, D.~A.~Vandenberg
  A\&A \textbf{329}, 87, (1998),
  
\bibitem{Carretta+2000}
  E.~Carretta, R.~G.~Gratton, G.~Clementini, F.~Fusi Pecci,
  ApJ \textbf{533}, 215, (2000),
  doi = {10.1086/308629}

\bibitem{Saunders+1990}
  W.~Saunders et al.
  MNRAS \textbf{242}, 318, (1990),
  doi = {10.1093/mnras/242.3.318}

\bibitem{Saunders+1991}
  W.~Saunders et al.
  Nature \textbf{349}, 32, (1991),
  doi = {10.1038/349032a0}

\bibitem{Efstathiou+1990}
  G.~Efstathiou, W.~J.~Sutherland, S.~J.~Maddox,
  Nature \textbf{348}, 705, (1990),
  doi = {10.1038/348705a0}

\bibitem{Loveday+1992}
  J.~Loveday et al.,
  ApJ \textbf{400}, L43, (1992),
  doi = {10.1086/186645}

\bibitem{Bunn+1997}
  E.~F.~Bunn \& M.~White,
  ApJ \textbf{480}, 6, (1997),
  doi = {10.1086/303955}

\bibitem{Bond+1999}
  J.~R.~Bond \& A.~H.~Jaffe,
  Roy Soc London Phil Tr A \textbf{357},57, (1999),
  doi = {10.1098/rsta.1999.0314}

\bibitem{Peacock1997}
  J.~A.~Peacock,
  MNRAS \textbf{284},885, (1997),
  doi = {10.1093/mnras/284.4.885}

\bibitem{Freedman+2001}
  W.~L.~Freedman et al.,
  ApJ \textbf{553}, 47, (2001)
  doi = {10.1086/320638}

\bibitem{Kelson+1996}
  D.~D.~Kelson et al.,
  ApJ \textbf{463}, 26, (1996)
  doi = {10.1086/177221}

\bibitem{Ferrarese+2000}
  L.~Ferrarese et al.,
  ApJS \textbf{128}, 431, (2000)
  doi = {10.1086/313391}

\bibitem{Gibson+2000}
  B.~K.~Gibson et al.,
  ApJ \textbf{529}, 723, (2000)
  doi = {10.1086/308306}

%\bibitem{Saha+1996a}
%  A.~Saha, A.~R.~Sandage et al.,
%  ApJ \textbf{466}, 55, (1996a)
%  doi = {10.1086/177493}

\bibitem{Saha+1994}
  A.~Saha et al.
  ApJ \textbf{425}, 14, (1994)
  doi = {10.1086/173957}

\bibitem{Freedman+1994a}
  W.~L.~Freedman et al.
  ApJ \textbf{427}, 628, (1994a)
  doi = {10.1086/174172}

\bibitem{Freedman+1994b}
  W.~L.~Freedman et al.
  Nature \textbf{371}, 757, (1994b)
  doi = {10.1038/371757a0}

\bibitem{Turner+1998}
  A.~Turner et al.
  ApJ \textbf{505}, 207, (1998)
  doi = {10.1086/306150}

\bibitem{Saha+1996}
  A.~Saha, A.~R.~Sandage et al.,
  ApJS \textbf{107}, 693, (1996)
  doi = {10.1086/192378}

\bibitem{Saha+1997}
  A.~Saha, A.~R.~Sandage et al.,
  ApJ \textbf{486}, 1, (1997)
  doi = {10.1086/304507}

\bibitem{Saha+1999}
  A.~Saha, A.~R.~Sandage et al.,
  ApJ \textbf{522}, 802, (1999)
  doi = {10.1086/307693}

\bibitem{Saha+2001}
  A.~Saha, A.~R.~Sandage et al.,
  ApJ \textbf{562}, 314, (2001)
  doi = {10.1086/323529}

\bibitem{Sandage+1996a}
  A.~R.~Sandage, A.~Saha et al.,
  ApJ \textbf{460}, L15, (1996a)
  doi = {10.1086/309973}

\bibitem{Tammann+2003}
  G.~A.~Tammann, A.~R.~Sandage, A.~Saha et al.,
  Proc. HST Symp. \textbf{14}, 222, (2003)
  doi = {10.1086/309973}

\bibitem{Sandage+1996b}
  A.~R.~Sandage \& G.~A.~Tammann,
  ApJ \textbf{464}, L51, (1996b)
  doi = {10.1086/310083}

\bibitem{Sandage1996a}
  A.~R.~Sandage,
  AJ \textbf{111}, 1, (1996a)
  doi = {10.1086/117755}

\bibitem{Sandage1996b}
  A.~R.~Sandage,
  AJ \textbf{111}, 18, (1996b)
  doi = {10.1086/117756}

\bibitem{Federspiel+1998}
  M.~Federspiel, G.~A.~Tammann, A.~R.~Sandage,
  ApJ \textbf{495}, 115, (1998)
  doi = {10.1086/305263}

\bibitem{Phillips1993}
  M.~M.~Phillips,
  ApJ \textbf{413}, L105, (1993)
  doi = {10.1086/186970}

\bibitem{Jha+1999}
  S.~Jha et al.,
  ApJS \textbf{125}, 73, (1999)
  doi = {10.1086/313275}

\bibitem{Phillips+1999}
  M.~M.~Phillips et al.,
  AJ \textbf{118}, 1766, (1999)
  doi = {10.1086/301032}

\bibitem{Wang2000}
  Y.~Wang,
  ApJ \textbf{536}, 531, (2000)
  doi = {10.1086/308958}

\bibitem{Hernandez+2000}
  M.~Hernandez et al.,
  MNRAS \textbf{319}, 223, (2000)
  doi = {10.1046/j.1365-8711.2000.03841.x}

\bibitem{Schmidt+1998}
  B.~P.~Schmidt et al.,
  ApJ \textbf{507}, 46, (1998)
  doi = {10.1086/306308}

\bibitem{Perlmutter+1999}
  S.~Perlmutter et al.,
  ApJ \textbf{517}, 565, (1999)
  doi = {10.1086/307221}

\bibitem{Smoot+1992}
  G.~F.~Smoot et al.,
  ApJ \textbf{396}, L1, (1992)
  doi = {10.1086/186504}

\bibitem{Bennett+1996}
  C.~L.~Bennett et al.,
  ApJ \textbf{464}, L1, (1996)
  doi = {10.1086/310075}

\bibitem{Hanany+2000}
  S.~Hanany et al.,
  ApJ \textbf{545}, L5, (2000)
  doi = {10.1086/317322}

\bibitem{deBernardis+2000}
  P.~de Bernardis et al.,
  Nature \textbf{404}, 955, (2000)
  doi = {10.1038/35010035}

\bibitem{Netterfield+2002}
  C.~B.~Netterfield et al.,
  ApJ \textbf{571}, 604, (2002)
  doi = {10.1086/340118}

\bibitem{Bartlett+1995}
  J.~G.~Bartlett et al.,
  Science \textbf{267}, 980, (1995)
  doi = {10.1126/science.267.5200.980}

\bibitem{White+1995}
  M.~White et al.,
  MNRAS \textbf{276}, L69, (1995)
  doi = {10.1093/mnras/276.1.L69}

\bibitem{Lineweaver+1997}
  C.~H.~Lineweaver et al.,
  A\&A \textbf{322}, 365, (1997)
  doi = {10.48550/arXiv.astro-ph/9610133}

\bibitem{Willick1994}
  J.~A.~Willick,
  ApJS \textbf{92}, 1, (1994)
  doi = {10.1086/191957}

\bibitem{Willick+1995}
  J.~A.~Willick et al.
  ApJ \textbf{446}, 12, (1995)
  doi = {10.1086/175762}

\bibitem{Willick+1996}
  J.~A.~Willick et al.
  ApJ \textbf{457}, 460, (1996)
  doi = {10.1086/176746}

\bibitem{Willick+1997a}
  J.~A.~Willick et al.
  ApJS \textbf{109}, 333, (1997a)
  doi = {10.1086/312983}

\bibitem{Willick+1997b}
  J.~A.~Willick et al.
  ApJ \textbf{486}, 629, (1997b)
  doi = {10.1086/304551}

\bibitem{Willick+1998}
  J.~A.~Willick, \& M.~A.~Strauss,
  ApJ \textbf{507}, 64, (1998)
  doi = {10.1086/306314}

\bibitem{Willick+2001}
  J.~A.~Willick, \& P.~Batra,
  ApJ \textbf{548}, 564, (2001)
  doi = {10.1086/319005}

\bibitem{Strauss+1995}
  M.~A.~Strauss \& J.~A.~Willick,
  PhysReports \textbf{261}, 271, (1995)
  doi = {10.1016/0370-1573(95)00013-7}

\bibitem{Courteau+2000}
  S.~Courteau \& J.~A.~Willick,
  ASP Conf. Ser. \textbf{201}, (2000)
  
\bibitem{Hudson1993}
  M.~J.~Hudson,
  MNRAS \textbf{265}, 43, (1993)
  doi = {10.1093/mnras/265.1.43}
  
\bibitem{Hudson1994}
  M.~J.~Hudson,
  MNRAS \textbf{266}, 475, (1994)
  doi = {10.1093/mnras/266.2.475}
  
%\bibitem{Hudson+1995}
%  M.~J.~Hudson et al.,
%  MNRAS \textbf{274}, 305, (1995)
%  doi = {10.1093/mnras/274.1.305}
  
\bibitem{Dekel+1990}
  A.~Dekel, E.~Bertschinger, S.~M.~Faber,  
  ApJ \textbf{364}, 349, (1990)
  doi = {10.1086/169418}
  
\bibitem{Dekel+1993}
  A.~Dekel et al.,
  ApJ \textbf{412}, 1, (1993)
  doi = {10.1086/172896}
  
\bibitem{Dekel+1999}
  A.~Dekel et al.,
  ApJ \textbf{522}, 1, (1999)
  doi = {10.1086/307636}
  
\bibitem{Giovanelli+1997}
  R.~Giovanelli et al.,
  ApJ \textbf{477}, L1, (1997)
  doi = {10.1086/310521}
  
\bibitem{Freudling+1999}
  W.~Freudling et al.,
  ApJ \textbf{523}, 1, (1999)
  doi = {10.1086/307707}

\bibitem{Pierce+1988}
  M.~J.~Pierce \& R.~B.~Tully,
  ApJ \textbf{330}, 579, (1988)
  doi = {10.1086/307707}

\bibitem{Tully+2000}
  R.~B.~Tully \& M.~J.~Pierce,
  ApJ \textbf{533}, 744, (2000)
  doi = {10.1086/308700}

\bibitem{Shaya+1995}
  E.~J.~Shaya, P.~J.~E.~Peebles, R.~B.~Tully, 
  ApJ \textbf{454}, 15, (1995)
  doi = {10.1086/176460}

\bibitem{Tonry+1981}
  J.~L.~Tonry \& M.~Davis,
  ApJ \textbf{246}, 680, (1981)
  doi = {10.1086/158965}

\bibitem{Karachentsev+2002}
  I.~D.~Karachentsev et al.
  A\&A \textbf{389}, 812, (2002)
  doi = {10.1051/0004-6361:20020649}

\bibitem{Sakai+2000}
  S.~Sakai et al.
  ApJ \textbf{529}, 698, (2000)
  doi = {10.1086/308305}

\bibitem{Efstathiou2020}
  G.~Efstathiou,
  arXiv:2007.10716v2, (2020)
  
\bibitem{Efstathiou+2021}
  G.~Efstathiou \& S.~Gratton,
  QJAp \textbf{4}, 8, (2021)
  doi = {10.21105/astro.1910.00483}

\bibitem{Planck2020}
  Planck Collaboration et al.,
  A\&A \textbf{641}, A6, (2020)
  doi = {10.1051/0004-6361/201833910}
  
\bibitem{Planck2014}
  Planck Collaboration et al.,
  A\&A \textbf{571}, A16, (2014)
  doi = {10.1051/0004-6361/201321591}
  
\bibitem{Aubourg+2015}
  E.~Aubourg et al.,
  PhRvD \textbf{92}, 123516, (2015)
  doi = {10.1103/PhysRevD.92.123516}
  
\bibitem{Macaulay+2019}
  E.~Macaulay et al.,
  MNRAS \textbf{486}, 2184, (2019)
  doi = {10.1093/mnras/stz978}
  
\bibitem{Lemos+2019}
  P.~Lemos et al.
  MNRAS \textbf{483}, 4803, (2019)
  doi = {10.1093/mnras/sty3082}
  
\bibitem{dAmico+2020}
  G.~d'Amico et al.,
  JCAP \textbf{5}, 005, (2020)
  doi = {10.1088/1475-7516/2020/05/005}
  
\bibitem{Riess+2018}
  A.~G.~Riess et al.,
  ApJ \textbf{855}, 136, (2018)
  doi = {10.3847/1538-4357/aaadb7}
  
\bibitem{Riess+2022a}
  A.~G.~Riess et al.,
  ApJ \textbf{934}, L7, (2022)
  doi = {10.3847/2041-8213/ac5c5b}
  
\bibitem{Rizzi+2007}
  A.~G.~Rizzi et al.,
  ApJ \textbf{661}, 815, (2007)
  doi = {10.1086/516566}
  
\bibitem{Anand+2022}
  G.~S.~Anand et al.,
  ApJ \textbf{932}, 15, (2022)
  doi = {10.3847/1538-4357/ac68df}
  
\bibitem{Freedman+2020}
  W.~L.~Freedman et al.,
  ApJ \textbf{891}, 57, (2020)
  doi = {10.3847/1538-4357/ab7339}
  
\bibitem{Tully+2023}
  R.~B.~Tully et al.,
  ApJ \textbf{944}, 94, (2023)
  doi = {10.3847/1538-4357/ac94d8}

\bibitem{Blakeslee+2021}
  J.~P.~Blakeslee et al.,
  ApJ \textbf{911}, 65, (2021)
  doi = {10.3847/1538-4357/abe86a}

\bibitem{deJaeger+2020}
  T.~de Jaeger et al.,
  MNRAS \textbf{496}, 3402, (2020)
  doi = {10.1093/mnras/staa1801}

\bibitem{Howlett+2022}
  C.~Howlett et al.,
  MNRAS \textbf{515}, 953, (2022)
  doi = {10.1093/mnras/stac1681}

\bibitem{Kourkchi+2022}
  E.~Kourkchi et al.,
  MNRAS \textbf{511}, 6160, (2022)
  doi = {10.1093/mnras/stac303}

\bibitem{Springob+2007}
  C.~M.~Springob et al.,
  ApJS \textbf{172}, 599, (2007)
  doi = {10.1086/519527}

\bibitem{Springob+2014}
  C.~M.~Springob et al.,
  MNRAS \textbf{445}, 2677, (2014)
  doi = {10.1093/mnras/stu1743}

\bibitem{Qin+2018}
  F.~Qin et al.,
  MNRAS \textbf{477}, 5150, (2018),
  doi = {10.1093/mnras/sty928}

\bibitem{Hong+2019}
  T.~Hong et al.,
  MNRAS \textbf{487}, 2061, (2019),
  doi = {10.1093/mnras/stz1413}

\bibitem{McGaugh+2000}
  S.~S.~McGaugh et al.,
  ApJ \textbf{533}, L99, (2000),
  doi = {10.1086/312628}

\bibitem{Lelli+2016}
  F.~Lelli, S.~S.~McGaugh, J.~M.~Schombert,
  ApJ \textbf{816}, L14, (2016),
  doi = {10.3847/2041-8205/816/1/L14}

\bibitem{Haynes+2018}
  M.~P.~Haynes et al.,
  ApJ \textbf{861}, 49, (2018),
  doi = {10.3847/1538-4357/aac956}

\bibitem{Courtois+2015}
  H.~M.~Courtois \& R.~B.~Tully,
  MNRAS \textbf{447}, 1531, (2015),
  doi = {10.1093/mnras/stu2405}

\bibitem{Tully+2008}
  R.~B.~Tully et al.,
  ApJ \textbf{676}, 184, (2008),
  doi = {10.1086/527428}

\bibitem{Tully+2013}
  R.~B.~Tully et al.,
  AJ \textbf{146}, 86, (2013),
  doi = {10.1088/0004-6256/146/4/86}

\bibitem{Tully+2016}
  R.~B.~Tully, H.~M.~Courtois, J.~G.~Sorce,
  AJ \textbf{152}, 50, (2016),
  doi = {10.3847/0004-6256/152/2/50}
    
\bibitem{Tully2015}
  R.~B.~Tully, 
  AJ \textbf{149}, 171, (2015),
  doi = {10.1088/0004-6256/149/5/171}
    
\bibitem{Kourkchi+2017}
  E.~Kourkchi \& R.~B.~Tully, 
  ApJ \textbf{843}, 16, (2017),
  doi = {10.3847/1538-4357/aa76db}
    
\bibitem{Tempel+2017}
  E.~Tempel et al.,
  A\&A \textbf{602}, A100, (2017),
  doi = {10.1051/0004-6361/201730499}
    
\bibitem{Hayden+2013}
  B.~T.~Hayden et al.,
  ApJ \textbf{764}, 191, (2013),
  doi = {10.1088/0004-637X/764/2/191}
    
\bibitem{Childress+2013}
  M.~Childress et al.,
  ApJ \textbf{770}, 108, (2013),
  doi = {10.1088/0004-637X/770/2/108}
    
\bibitem{Roman+2018}
  M.~Roman et al.,
  A\&A \textbf{615}, A68, (2018),
  doi = {10.1051/0004-6361/201731425}
    
\bibitem{Reid+2019}
  M.~J.~Reid, D.~W.~Pesce, A.~G.~Riess,
  ApJ \textbf{886}, L27, (2019),
  doi = {10.3847/2041-8213/ab552d}
    
\bibitem{Pesce+2020}
  D.~W.~Pesce et al,.
  ApJ \textbf{891}, L1, (2020),
  doi = {10.3847/2041-8213/ab75f0}
    
\bibitem{Pietrzynski+2019}
  G.~Pietrzy\'nski et al,.
  ApJ \textbf{891}, L1, (2019),
  doi = {10.1038/s41586-019-0999-4}
    
\bibitem{Birrer+2022}
  S.~Birrer et al,.
  arXiv:2210.10833, (2022),
  doi = {10.48550/arXiv.2210.10833}
    
\bibitem{Abbott+2021}
  B.~P.~Abbott et al,.
  ApJ \textbf{909}, 218, (2021),
  DOI = {10.3847/1538-4357/abdcb7}
    
\bibitem{Moresco+2018}
  M.~Moresco et al,.
  ApJ \textbf{868}, 84, (2018),
  DOI = {10.3847/1538-4357/aae829}
    
\bibitem{Kozmanyan+2019}
  A.~Kozmanyan et al,.
  A\&A \textbf{621}, A34, (2019),
  DOI = {10.1051/0004-6361/201833879}
    
\bibitem{Alam+2021}
  S.~Alam et al,.
  PhRvD \textbf{103}, 083533, (2021),
  DOI = {10.1103/PhysRevD.103.083533}
    
  
\end{thebibliography}
\end{document}